\documentclass[reprint,amsmath,amssymb,aip,]{revtex4-1}
\usepackage{amsmath}
\usepackage{booktabs}
\usepackage{multibib}
\usepackage{braket}
\usepackage{amsfonts}
\usepackage{graphicx}
\usepackage{dcolumn}
\usepackage{url}
\usepackage{bm}
\usepackage[colorlinks=true,citecolor=blue,linkcolor=blue,urlcolor=blue]{hyperref}
\usepackage{setspace}

\begin{document}

\title{ Strained BiFeO$_{3}$ perovskite structure for enhanced photovoltaic effect.}

\author{Tewodros Eyob}
\email{tewodros.eyob@aau.edu.et}
\affiliation{Department of
Physics, Addis Ababa University, P.O. Box 1176, Addis Ababa, Ethiopia}

\author{Kenate Nemera}
\email{kenate.nemera@gmail.com}
\affiliation{Department of
Physics, Addis Ababa University, P.O. Box 1176, Addis Ababa, Ethiopia}

\author{Lemi Demeyu}
\email{lemi.demeyu@aau.edu.et}
\affiliation{Department of
Physics, Addis Ababa University, P.O. Box 1176, Addis Ababa, Ethiopia}
\date{\today}
\begin{abstract}
BiFeO$_{3}$ is multiferroic material with space group Pbnm exhibits coupling of both magnetic and electric orders under strain force. To analyze band gap Tauc plot extrapolation method is used  $\&$  remarkably smaller than some reported  literature values for space group R3c structure. The dielectric function $\varepsilon(\omega)$  demonstrated  that light energy upon transition  through BiFeO$_{3}$ has shown  a decline inconsistent pattern with index of refraction variation in range of 13 to 8, 8 to 5, 5 to 3, $\&$ 3 to 1  as per photon energy. Moreover, non crossing degenerate energy level and hybridized spin-orbit interaction lead to linear dispersion relation which is typical photonic nature with zero DM interaction. Thus, BiFeO$_{3}$ could  exhibit photonic property under strain force.  Generally strained structure had  enhanced photovoltaic effect with smaller optical band gap  because of $Dzyaloshinskii$-$Moriya(DM)$  and spin-orbit coupling interactions.

\end{abstract}

\maketitle


\section{INTRODUCTION}

Understanding and dealing crystallographic structure has been center of new material research findings,  exploring chemical and physical properties of materials with varies thermodynamics variables have been research interest for the last three decades \cite{1}. Howard et al\cite{2} has applied group-theoretical method to the space group and crystal structures of centrosymmetric, zone-boundary-tilted members of perovskite and elpasolite groups in relationships with bond lengths, bond angles, polyhedron volumes and distortions\cite{3}. Knight et al\cite{4} explained that KCa$F_{3}$ under go structural phase transitions at 560 k from a rhombohedral space group Cmcm to an orthorhombic Pbnm upon cooling. Lufaso and woodward showed KCa$F_{3}$ frequently crystallize as space-group Pbnm at ambient temperature\cite{5}. Comley noted that in SrTiO$_{3}$, rotating TiO$_{6}$ octahedron at corner of crystal system along any crystallographic directions result in irreducible representations\cite{6}. Obviously, under varies external variables crystal structure prefer to retain a simple  irreducible symmetric crystal structure. Bismuth ferrite exhibit varies nature and order either in bulk form or thin film over ranges of phase transformation temperature. 

\begin{figure*}[htp!]
\centering
\includegraphics[width= 5.50in]{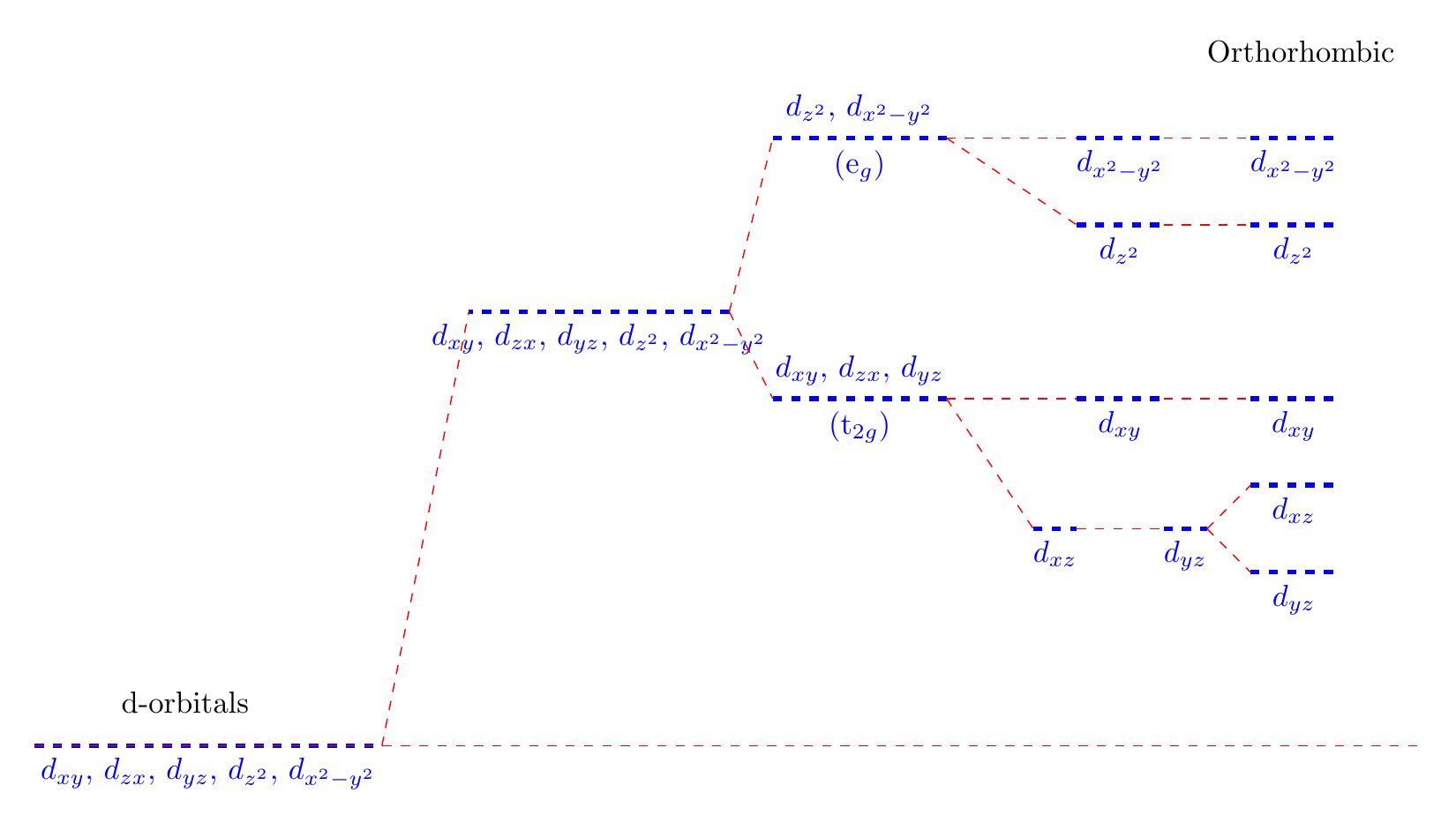}
\caption{\label{akqi} Crstal-field splitting and degenerate  energy levels of the Fe cation's d-states in a FeO$_{6}$ octahedron of orthorhombic distortions.}
\end{figure*}

Any BiFeO$_{3}$ distortion from the space group Pbnm  pervskite structure will result in degeneracy between the $d$-states labeled $t_{2g}$(triplet) and $e_{g}$(doublet). Deviation has been due to $FeO_{6}$ octahedron tilt and altered infinitesimally\cite{7}. However, infinitesimal deviation had significant impact on magnetic, electronic and optical property of BiFeO$_{3}$.
These degenerate energy levels ordered and filled by electrons based on Hund's rules: electron's spins must have highest multiplicity of both total spin $\&$ total angular momentum where start filling from lowest enegy level\cite{8}(see Fig.\ref{akqi}).

Crystal field theory model describes the magnetic behavior of crystallographic structure and significance of deviation angle which play critical role in an antisymmetric magnetic interaction, the so-called $Dzyaloshinskii$-$Moriya(DM)$, which is due to spin-orbit interaction $\&$ often induce a weak antiferromagnetism in ferrites\cite{9}.

Therefore, in this work we explore role and effect of strained antisymmetric BiFe$O_{3}$ crystal  structure which are derivative of Pbnm space group.

\begin{figure}[htp!]
\centering
\includegraphics[width= 3.50in]{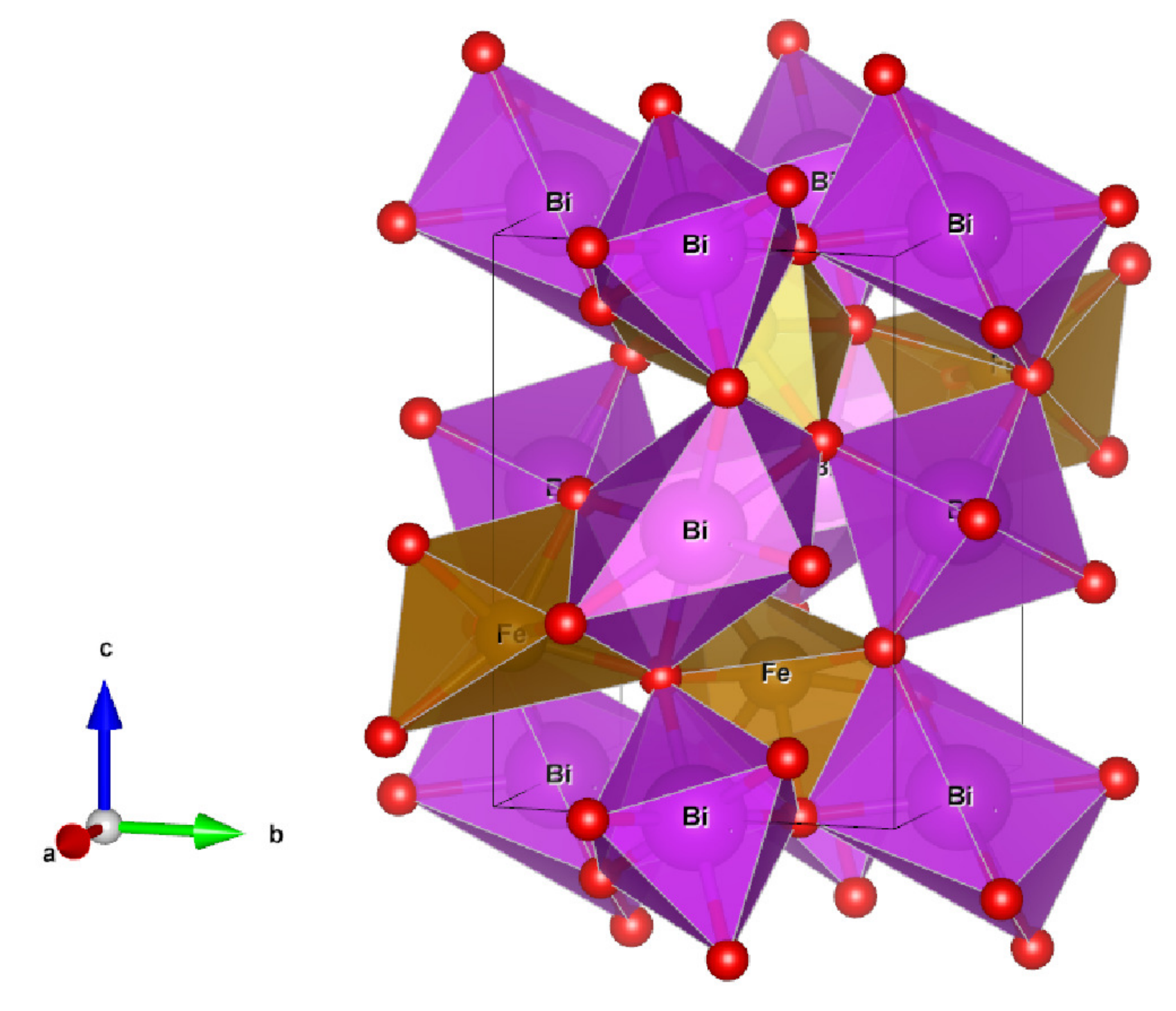}
\caption{\label{amj} Crystal structure of bulk BiFeO$_{3}$(Pbnm) illustrating the structural tilting of BiO$_{6}$ polyhedra relative to neighboring  polyhedrons resulting in the breaking of spatial inversion symmetry.}
\end{figure}

BiFeO$_{3}$ shows $\beta$ and $\gamma$ paraelectric orthorhombic phases between 1200K and 1100K temperatures with space group Pbnm structure\cite{10, 11}. The Pbnm structure made from simple cubic perovskite structure of BiFeO$_{3}$ unit cell with A-site cation at the corners and the B-site at the corner coordinated by six oxygen located at the face centers forming an oxygen octahedron, rotating the oxygen octahedron about three cubic axis $\bf{a}$, $\bf{b}$, and $\bf{c}$ by the angles $\bf{\alpha}$, $\bf{\beta}$, and $\bf{\gamma}$, respectively, neighboring  octahedron must be rotated as well, which minimize the symmetry\cite{12}. Then, structure of pseudo cubic formed with constituent unit cell of 12 oxygen anions, 4 B-site cations, and 4 A-site cations(see Fig.\ref{amj}).

In this work  we considered psuedo cubic structure  strained to reduced volume from original unstrained Pbnm structure as displayed on Table.\ref{table:non2lin}. 

Strain force vector directions are $\vec{a} \vec{b}$, $\vec{a} \vec{b} \vec{c}$, $\vec{a} \vec{c}$ and $\vec{b} \vec{c}$. Further Compressed to 15$^{\circ}$ between plane $\angle BC$ with $\bf{\alpha}$-angle for  amplified antisymmetric effect on Pbnm structure.

\section{Computational method}

$\emph{Ab initio}$ calculation were performed using the projector augmented wave(PAW) method as implemented in 
gpaw code\cite{13}. The electron wave-function is
approximated using the implementation of a projector augmented wave method\cite{14} $\&$ the Schr$\ddot{o}$dinger equation applied to total energy functions are solved self-consistently using the Kohn-Sham scheme \cite{15}. All electron waves are expanded over periodic potential plane with a method of P.E.Bl$\ddot{o}$chl\cite{16} having a band, a k-points, and a reciprocal lattice grid indices. A plane wave cut-off energy used is 520 eV. The sum of plane wave kinettic energy limit $\frac{1}{2}|G+E|^{2}<E_{Cutoff}$. The minimal energy $E_{Cutoff}$ where the plane wave is expanded in the reciprocal lattice vector grid G of frist Brillouin zone. The k-points within Brillouin zone (BZ) is chosen according to Monkhorst-Pack scheme \cite{17}, where a $\bf{K}$-meshes of 8x8x8 is used.

The interactions of the valence electrons with the core electrons and nuclei is treated using compensation charge both long-range electrostatics  and nodal shape of orbitals correctly addressed within a projector augmented wave (paw) data sets\cite{16, 18}.

The number of valence electrons considered for each element within the paw data sets is Bi(10d$^{10}$ 6s$^{2}$ 6p$^{3}$), Fe(3s$^{2}$ 3d$^{6}$ 4s$^{2}$), and O(2s$^{2}$ 2p$^{4}$). Geometry relaxations are carried out using BFGS minimizer \cite{19}, where optimizations of the atomic coordinates and the unit cell degrees of freedom is done within the concept of the Hellmann-Feynman forces and stresses \cite{20,21} calculated on the Born-Oppenheimer(BO) surface \cite{22}. The convergence criteria for the forces were set at 0.05 $eV/\AA$. The exchange-correlation energies are approximated within the generalized gradient approximation of PBE \cite{23} $\&$ a $\bf{k}$-mesh of 3x3x3 is used in a geometry relaxation calculations. Hubard model\cite{24} selected for better description of d-orbital of Fe : U$_{eff}$=7.5eV with spin polarized calculation applied.

In calculation of the band gap, we have used Tauc plot 
method\cite{25} which extrapolate by fitting a straight line to the linear segment to intersect  $\hbar \omega$-axis from the plot of $(\hbar \omega \alpha)^{2}$ against $\hbar \omega$.
\begin{equation}
(\hbar\omega \alpha)^{1/2}=A(\hbar\omega-E_{g})
\end{equation}
where $\hbar$ is the Plank's constant, $\omega$ is frequency , $\alpha$ is absorption coefficient, $E_{g}$ is band gap and A is proportionality constant.

For band structure calculation a special set of k-points mesh  8x8x8 is used and implemented exchange-correlation functionals within PBE approximation\cite{23}. 
   
Crystal field theory explains that degenerate d-orbitals of orthorhombic crystal structure  split into $t_{2g}$ triplet and $e_{g}$ doublet states due to the crystal field splitting energy. The Fe$^{+3}$ ion is surrounded by an octahedron oxygen with valance electrons of 5 d orbitals having spins aligned parallel producing 5 $\mu_{\beta}$ per Fe$^{+3}$ ion\cite{26}. The d orbitals barely interact with nearby Fe$^{+3}$ sites. However, exchange interaction made with six neighboring oxygen 2p orbitals. The general Hamiltonian becomes,
\begin{equation}
\begin{array}{cc} H=J_{AF}\sum_{\langle ij \rangle}S_{i}.S_{j}-J_{H}\sum_{i}s_{i}.S_{i}& \\
-\sum_{\langle ij \rangle, \sigma, \alpha, \beta} t_{i,j}^{\alpha, \beta}c^{\dagger}_{i, \alpha, \beta, \sigma} c_{j, \beta, \sigma}+D_{ij}.(\vec{S_{i}}X\vec{S_{j}})& \\
\end{array}
\label{maspin}
\end{equation}
where $\alpha$, $\beta$ stands for d-orbitals of anisotropic hopping matrix elements $t_{ij}^{\alpha \beta}$ with several possible outcomes of creation operator with introduction of canted ant-ferro magnetism, $J_{AF}$ is antiferromagnetic super exchange coupling constant between spins($S_{i}.S_{j}$) of $t_{2g}$ and $e_{g}$ energy levels and $J_{H}$ is Hund's coupling constant on-site spins($S_{i}.S_{i}$) and the $DM$ interaction\cite{27}. Eq.(\ref{maspin}) show that the source of multiferroic order are coupling constant and the DM interactions.

One can analyze and understand  material's optical response property once incident photon energy imparted to  the electron, $\&$ material's response is  given as
\begin{equation}
\varepsilon(\omega)=\varepsilon_{1}(\omega)+i\varepsilon_{2}(\omega)
\end{equation}
The imaginary part $\varepsilon_{2}(\omega)$ of dielectric function is calculated using the following expression 
\begin{equation}
\varepsilon_{2}(\omega)=\left(\frac{4\pi^{2}e^{2}}{m^{2}\omega^{2}}\right)\sum_{i,j}\int i|M|j^{2}f_{i}(1-f_{i})\delta(E_{f}-E_{i}-\omega)d^{3}k
\end{equation}

Where, M is the diplole matrix, i and j are initial and final states respectively; $f_{i}$ is the Fermi distribution function for the $i^{th}$ state and $\omega$ is the frequency of the incident photon.

Real part $\varepsilon_{1}(\omega)$ of the dielectric function can be found from Kramers-Kronig equation\cite{28}.

\begin{equation}
\varepsilon_{1}(\omega)=1+\frac{2}{\pi}P\int_{0}^{\infty}\frac{\omega^{'}\varepsilon_{2}(\omega^{'})d\omega^{'}}{\omega^{'2}-\omega^{2}}
\end{equation}
Where, P stands for the principle value of the integral.

From dielectric function, all the other optical properties, such as, reflectivity $R$. 
\begin{equation}
R(\omega)=\frac{(n-1)^{2}+\kappa^{2}}{(n-1)^{2}+\kappa^{2}}=\left|\frac{\sqrt{\varepsilon(\omega)+1}}{\sqrt{\varepsilon(\omega)+1}}\right|^{2}
\end{equation}
Refractive index $n_{0}$, extinction coefficient $\kappa$ of complex refractive index $\tilde{n}$ written as\cite{29,30}
\begin{equation}
\tilde{n}=n_{0}+i\kappa=\pm\sqrt{\tilde{\mu}_{r}\tilde{\varepsilon}_{r}}
\label{madf}
\end{equation}
where, $n_{0}$ is the real part of the complex refractive index (refractive index) and $\kappa$ is the imaginary part of the complex refractive index (extinction co-efficient), $\tilde{\mu}_{r}$ is magnetic response or relative permeability and $\tilde{\varepsilon}_{r}$ relative permittivity  of complex refractive index $\tilde{n}$, the magnetic response $\tilde{\mu}_{r}$ of most dielectric materials is negligible when $\tilde{\mu}_{r}$ considered unity. Here we used tetrahedron method  $\&$ proved  efficient espacially for magnetically induced dielectric function calculation\cite{31}.

\section{RESULTS AND DISCUSSION}
Pseudo cubic paraelectric orthorhombic structure with Pbnm space group with lattice parameters a=5.482 $\AA$, b=5.73 $\AA$ and c=7.86 $\AA$ has been considered for varies antisymmetric structure calculation. unstrained Pbnm space group is similar to experimental parameters\cite{32}.

\begin{figure}[htp!]
\includegraphics[width= 3.5in]{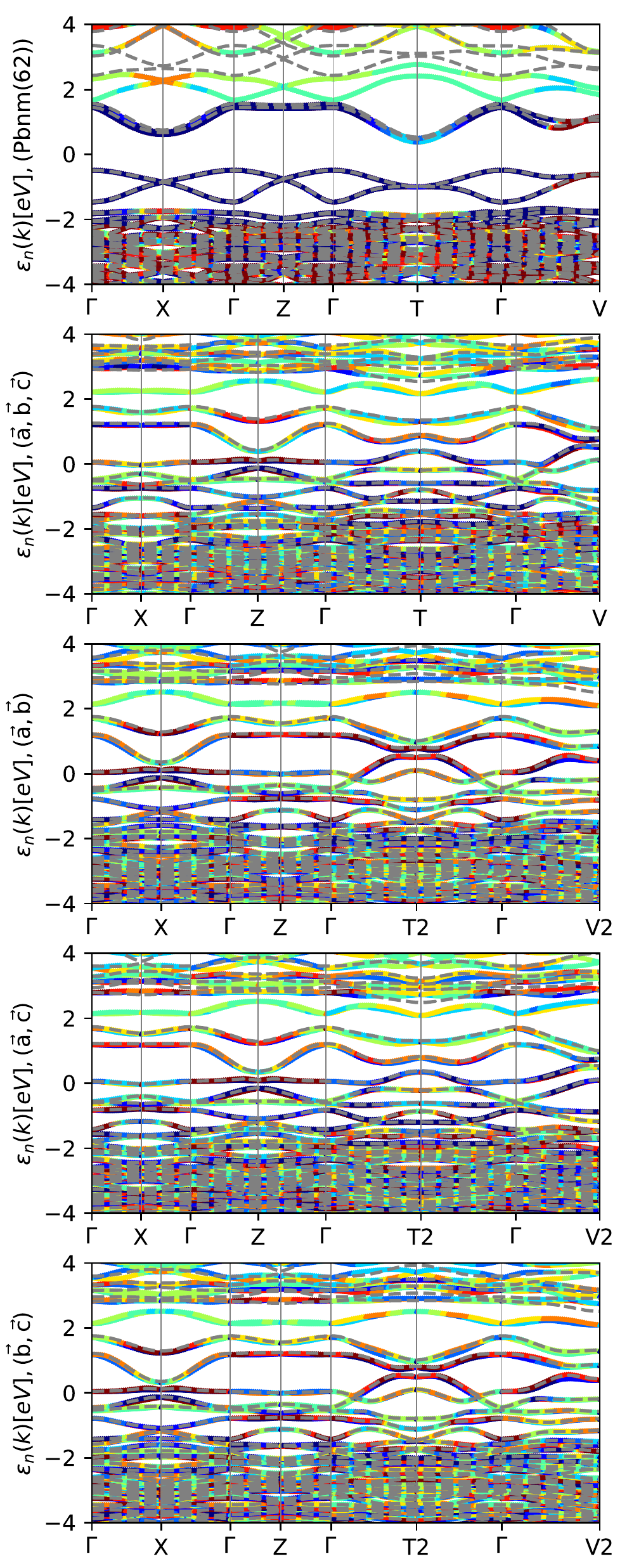}
\caption{\label{akqics} Scatter spin-orbit coupled band structure calculated using Eq.(\ref{maspin}).The colors show the spin character: blue designated as spin down, red as spin up and dashed gray lines as bands without spin-orbit coupling respectively(For interpretation of the references to color in this figure legend, the reader is referred to the web version of this article). }
\end{figure}

\begin{figure}[htp!]
\includegraphics[width= 3.5in]{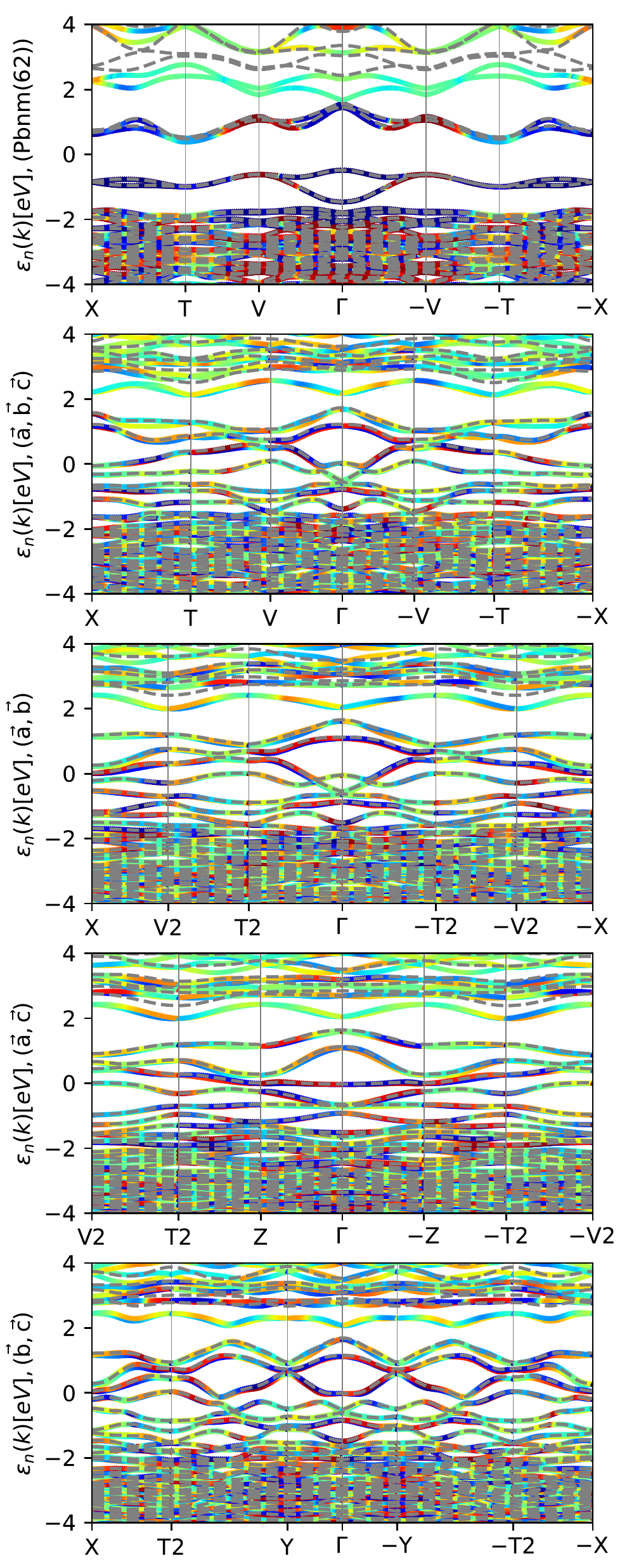}
\caption{\label{akqim} Time-reversal symmetry band structure scatter plot are either red or blue signaling that the bands are approximate eigenstates of the spin projection operator along the z-axis, spin up is displayed as red and spin down is displayed as blue(For interpretation of the references to color in this figure legend, the reader is referred to the web version of this article).}
\end{figure}

Band structure calculated in Fig.(\ref{akqics}) and Fig.(\ref{akqim}) based on spin-orbit coupling implemented in electronic structure code GPAW\cite{13}. One can input crystallographic  information file of any space group  for special k-point band paths $\&$ export from online website which made available at \url{http://www.materialscloud.org/tools/seekpath/}\cite{33}.

All selected antisymmetric crystal structure had simulatanously broken spaitial-inversion and time-reversal symmetery\cite{34} upon symmetry operation this in turn favors magnetic and electric order pairing. Therefore, degenerate bands at the boundary of the Brillouin zone splits up and narrows the gaps between conduction $\&$ valance bands near $\Gamma$ point.

Degenerate bands shown where there is band crossing except those strained in $\vec{a} \vec{c}$, $\vec{a} \vec{b}  \vec{c}$ and Pbnm structure.Two high energy degenerate levels($e_{g}$) and three low energy levels($t_{2g}$) not including two anti-bonding and bonding bands which were created due to hybridization of spin-orbit interaction either constructive or distractive way.

\begin{figure*}[htp!]
\centering
\includegraphics[width= 6.0in]{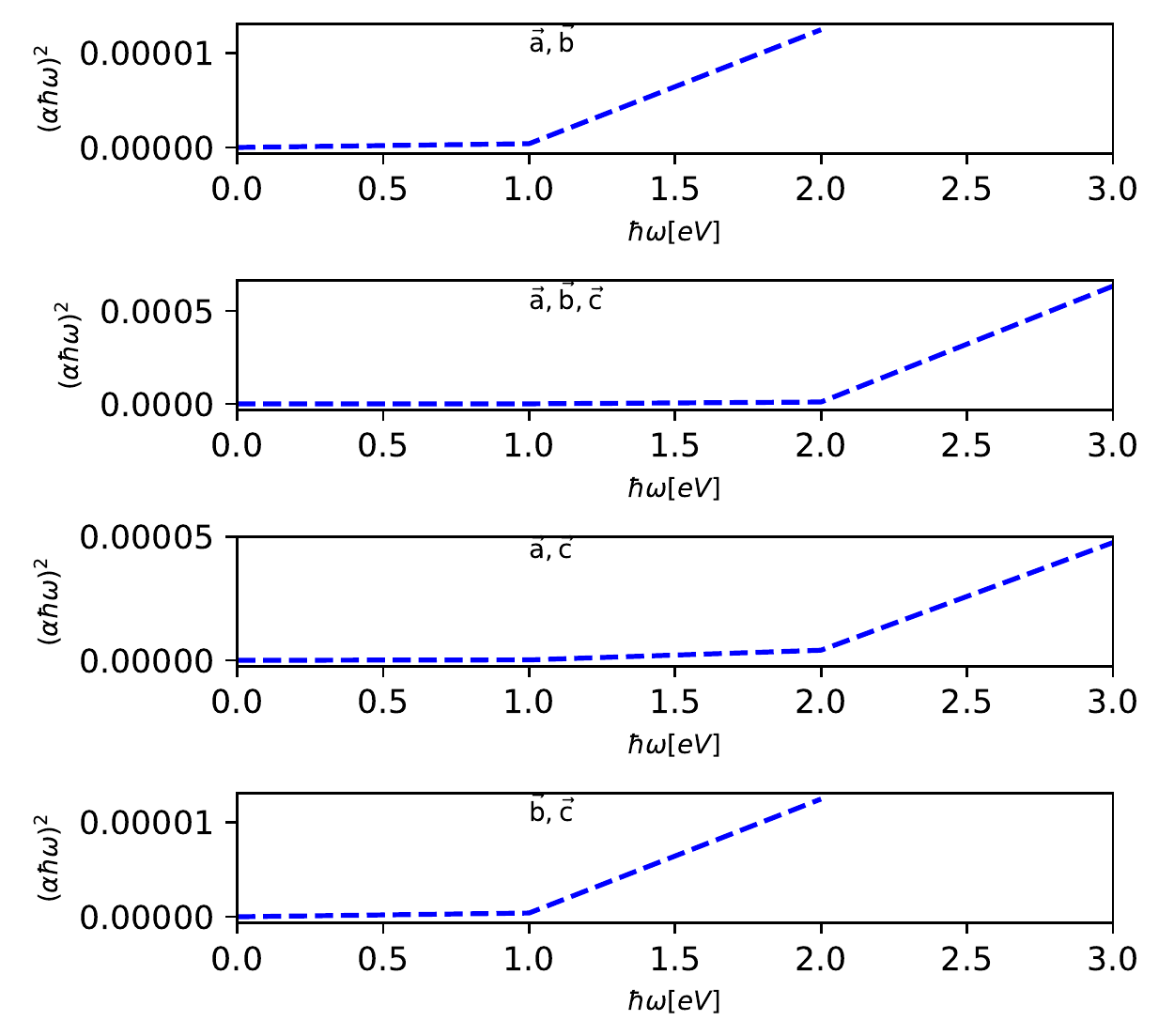}
\caption{\label{akqieg} Tauc plots vs. photon energy $\hbar \omega$ for direct band gap $E_{g}$ transition together with a linear extrapolation of $(\alpha \hbar \omega)^{2}$ for antisymmetric strain vector  direction $\vec{a} \vec{b}$, $\vec{a} \vec{b} \vec{c}$, $\vec{a} \vec{c}$, $\vec{b} \vec{c}$.}
\end{figure*}

As depicted on Table.{\ref{table:non2lin}} those structure strained on $\vec{a} \vec{c}$, $\vec{a} \vec{b}  \vec{c}$ resulted in simultaneous magnetic and electric order pairing  with larger optical band gap as shown in Fig.(\ref{akqieg}). However, only polarization exhibited in $\vec{a} \vec{c}$,  since strain vector direction or polarization is perpendicular to the spins,  magnitazition becomes zero due to Lebeugle \textit{et al}\cite{42} $\vec{P} \propto \vec{e}_{ij}X(\vec{S}_{i}X\vec{S}_{j})$ relation. Thus, DM interaction result in null. Consequently, reduced total energy encountered according to Eq.(\ref{maspin}) because of removal of magneto-electric coupling term, that is, $E_{DM}=(\vec{P}X\vec{e}_{ij}).(\vec{S}_{i}X\vec{S}_{j})$, where $S_{i,j}$ are local spins and $e_{ij}$ is the unit vector connecting the two sites.

\begin{figure}[htp!]
\includegraphics[width= 3.50in]{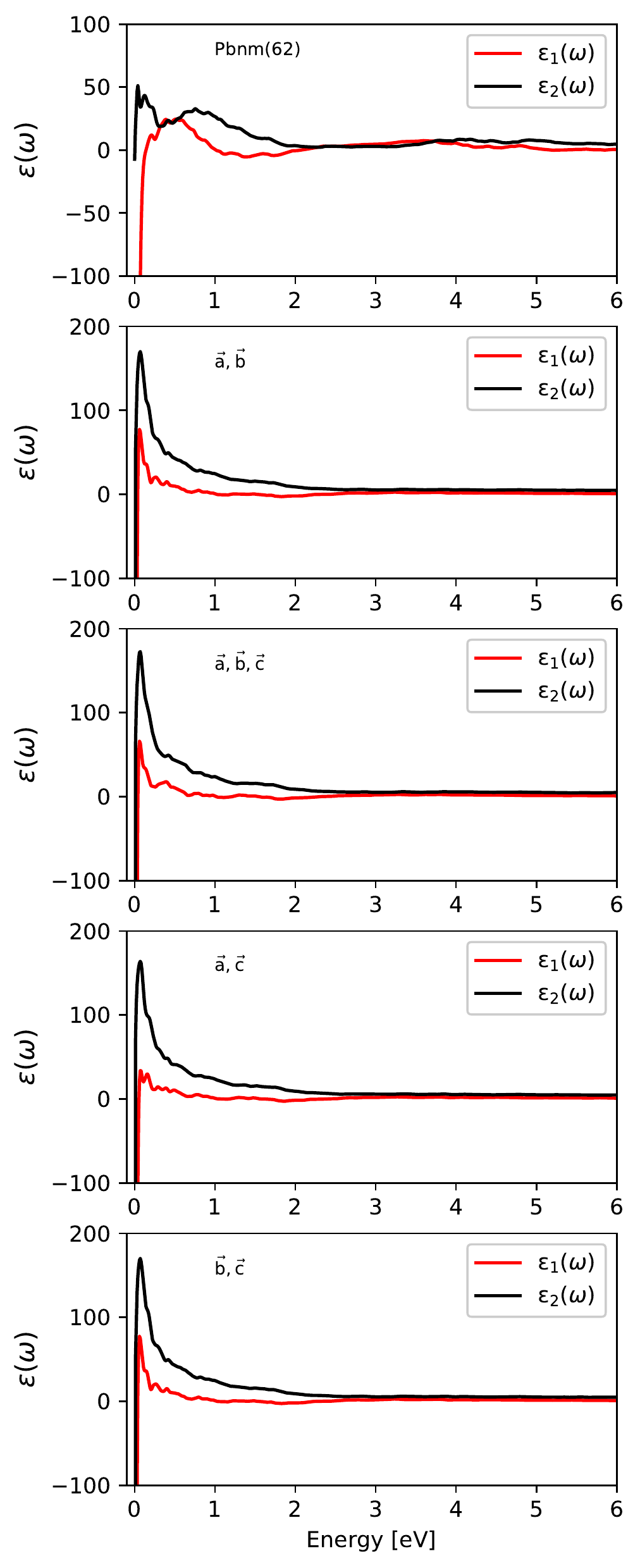}
\caption{\label{akqidf} Calculated dielectric function $\varepsilon(\omega)$ and $\tilde{n}(\omega)$ according to Eq.(\ref{madf}). Colors: real part $\varepsilon_{1}(\omega)$ is given in red color solid lines, imaginary part $\varepsilon_{2}(\omega)$ is given in black color solid lines, $\&$  $\tilde{n}(\omega)$ is given in blue color solid lines(For interpretation of the references to color in this figure legend, the reader is referred to the web version of this article).}
\end{figure}

Magnetic and electric order coupling shown in all antisymetric structures except in paraelectric Pbnm $\&$ $\vec{a} \vec{c}$ strained structures. However, in Fig.(\ref{akqics})  $\vec{a} \vec{c}$ strained behave like photonic crystal band structure\cite{35, 36, 37, 38, 39, 40} in such away that the dispersion relation is linear unlike electronic band structure which is parabolic, and spin-orbit interaction evolve with possible $S=1$ states contrast to electron $S=1/2$ state $\&$  ignored magnetization. Moreover, refractive index is relatively smallest as shown in Fig.(\ref{akqidf}) compared to others $\&$ decline with range of 13 to 8, 8 to 5, 5 to 3 $\&$ 3 to 1 as per incident photon energy. Hence, BiFeO$_{3}$ strained in specific direction could be considered as photonic crystal. Moreover, strain could promote remarkably fast photon energy transfer for photon's energy close to smaller optical band gap and enhanced photovoltaic effect with greater photo sensitivity.
\newcommand{\ra}[1]{\renewcommand{\arraystretch}{#1}}
\begin{table*}\centering \ra{1.3}
\caption{ Strain force acted on symmetric Pbnm space group  with possible strain vector direction resulting in varies morphological, magneto electric changes with respective band gap displayed for comparison.}
\begin{tabular}{@{}ccccccccccccc@{}}\toprule \hline\hline
 &&  &&  &&  && && \\ 
Space group  && Vector Strain Direction&& Unit-cell volume($\AA^{3}$) && Cell parameters && Bandgap(eV) && Magnetic Moment($\mu\mu_{B}$)\\ \hline 
\hline
&& &&  && $a=5.482234 \AA$ && && \\
&& &&  && $b=5.729905 \AA$ && &&\\
$Pbnm$&& - && 247.04087 &&$c=7.864368 \AA $&& 2.71\cite{41}&& 0.000\\
&& &&  && $\alpha=90^{\circ}$ && &&\\
&& &&  && $\beta=90^{\circ}$&& &&\\
&& &&  && $\gamma=90^{\circ}$ && &&\\
\hline
&& &&  && $a=5.48223 \AA$ &&  &&\\
&& &&  && $b=5.73078 \AA$ &&  &&\\
$P_{1}$&& $\vec{a}, \vec{b}$ && 247.09827602 &&$c=7.865 \AA $&& 0.915853&&178\\
&& &&  && $\alpha=74.0^{\circ}$ &&  &&\\
&& &&  && $\beta=89.2715^{\circ}$&& &&\\
&& &&  && $\gamma=90^{\circ}$ &&  &&\\
\hline 
&& &&  && $a=5.48223 \AA$ &&  &&\\  
&& &&  && $b=5.73078 \AA$ && &&\\
$P_{1}$&&$\vec{a}, \vec{b}, \vec{c}$ && 247.09827602 &&$c=7.865 \AA $&& 1.93949&&44\\
&& &&  && $\alpha=75.0^{\circ}$ &&  &&\\
&& &&  && $\beta=89.2715^{\circ}$&&  &&\\
&& &&  && $\gamma=89.0002^{\circ}$ &&  &&\\
\hline
&& &&  && $a=5.48223 \AA$ &&  &&\\
&& &&  && $b=5.73078 \AA$ &&  &&\\
$P_{1}$&&$\vec{a}, \vec{c}$ && 247.09827602 &&$c=7.865 \AA $&& 1.86041&&0.000\\
&& &&  && $\alpha=74.0^{\circ}$ &&  &&\\
&& &&  && $\beta=90^{\circ}$&&  &&\\
&& &&  && $\gamma=89.0002^{\circ}$ &&  &&\\
\hline
&& &&  && $a=5.48223 \AA$ &&  &&\\
&& &&  && $b=5.73078 \AA$ &&  &&\\
$P_{1}$&&$\vec{b}, \vec{c}$ && 247.09827602 &&$c=7.865 \AA $&& 0.923963&&90\\
&& &&  && $\alpha=74.0^{\circ}$ && &&\\
&& &&  && $\beta==89.2715^{\circ}$&&  &&\\
&& &&  && $\gamma=90^{\circ}$ &&  &&\\
\hline\hline
\bottomrule
\end{tabular}
\label{table:non2lin}
\end{table*}

\begin{acknowledgments}
We are grateful to Ministry of Science and Higher Eduction of Ethiopia for financial support and International Science Program, Uppsala University, Sweden are gratefully acknowledged for providing computational facilities.  We are also thankful to the office of VPRTT of Addis Ababa university for supporting this research under a grant number AR/032/2021.
\end{acknowledgments}

\newpage
\section{Reference}
\bibliographystyle{aps-nameyear}      


\begin{thebibliography}{0}%
\makeatletter
\providecommand \@ifxundefined [1]{%
 \@ifx{#1\undefined}
}%
\providecommand \@ifnum [1]{%
 \ifnum #1\expandafter \@firstoftwo
 \else \expandafter \@secondoftwo
 \fi
}%
\providecommand \@ifx [1]{%
 \ifx #1\expandafter \@firstoftwo
 \else \expandafter \@secondoftwo
 \fi
}%
\providecommand \natexlab [1]{#1}%
\providecommand \enquote  [1]{``#1''}%
\providecommand \bibnamefont  [1]{#1}%
\providecommand \bibfnamefont [1]{#1}%
\providecommand \citenamefont [1]{#1}%
\providecommand \href@noop [0]{\@secondoftwo}%
\providecommand \href [0]{\begingroup \@sanitize@url \@href}%
\providecommand \@href[1]{\@@startlink{#1}\@@href}%
\providecommand \@@href[1]{\endgroup#1\@@endlink}%
\providecommand \@sanitize@url [0]{\catcode `\\12\catcode `\$12\catcode
  `\&12\catcode `\#12\catcode `\^12\catcode `\_12\catcode `\%12\relax}%
\providecommand \@@startlink[1]{}%
\providecommand \@@endlink[0]{}%
\providecommand \url  [0]{\begingroup\@sanitize@url \@url }%
\providecommand \@url [1]{\endgroup\@href {#1}{\urlprefix }}%
\providecommand \urlprefix  [0]{URL }%
\providecommand \Eprint [0]{\href }%
\providecommand \doibase [0]{http://dx.doi.org/}%
\providecommand \selectlanguage [0]{\@gobble}%
\providecommand \bibinfo  [0]{\@secondoftwo}%
\providecommand \bibfield  [0]{\@secondoftwo}%
\providecommand \translation [1]{[#1]}%
\providecommand \BibitemOpen [0]{}%
\providecommand \bibitemStop [0]{}%
\providecommand \bibitemNoStop [0]{.\EOS\space}%
\providecommand \EOS [0]{\spacefactor3000\relax}%
\providecommand \BibitemShut  [1]{\csname bibitem#1\endcsname}%
\let\auto@bib@innerbib\@empty
\end{thebibliography}%


\nocite{*}
\begin{thebibliography}{}

\bibitem{1} Howard, C.J, Luca, V. $\&$ Knight, K.S.(2001): High temperature phase transitions in tungsten trioxide-the last word?. J. Phys., Condensed Matter, $\bf{13}$, 1-11, \url{https://doi.org/10.1088/0953-8984/14/3/308}.

\bibitem{2}
Howard, C.J., Kennedy, B.J. $\&$ Woodward, P.M.(2003).
Ordered double perovskites- a group-theoretical analysis, Acta Crystallogr: $\bf{59}$, 463-471, \url{https://doi.org/10.1107/S0108768103010073}.

\bibitem{3} Knight, K.S(2009a). Parameterization of the crystal structure of centrosymmeteric zone-boundary-tilted perovskites: An analysis in terms of symmetery-adapted basis-vectors of the cubic aristotype phase Can. Mineral. $\bf{47}$, 381-400, \url{https://doi.org/10.3749/canmin.47.2.381}.

\bibitem{4} Knight, K.S, Darlington, C.N.W. $\&$ Wood, I.G.(2005b). The crystal structure of KCaF$_{3}$ at 4.2 and 300 K: a revaluation using high-resolution powder neutron diffraction, Powder Diffraction   $\bf{20}$, 7-13, \url{https://doi.org/10.1154/1.1835959}.

\bibitem{5} Lufaso, M.W., Woodward, P.M.(2001).Prediction of crystal structures of perovskites using the software SPuDS. Acta Crystallogr $\bf{57}$, 725-738, \url{http://doi.org/10.1107/S0108768101015282}.


\bibitem{6} Comley, R.A.(1964).Lattice dynamics and phase transitions of strontium titanate. Phys Rev. $\bf{134}$, A981-A997, \url{https://doi.org/10.1103/PhysRev.134.A981}.

\bibitem{7}M. Tokunaga, M. Akaki, T. Ito, S. Miyahara, A. Miyake, H. Kuwahara $\&$ N. Furukawa (2015). Magnetic control of transverse electric polarization in BiFeO${_3}$. Nat. Commun. $\bf{6}$:5878 \url{doi: 10.1038/ncomms6878}.


\bibitem{8}  John D. Morgan,and Werner Kutzelnigg(1993). Hund's Rules, the Alternating Rule, and Symmetry Holes. J. Phys. Chem. $\bf{97}$, 2425-2434, \url{https://doi.org/10.1021/j100112a051}.

\bibitem{9} Cheong, SW., Mostovoy, M. Multiferroics(2007) a magnetic twist for ferroelectricity. Nature Mater $\bf{6}$, 13-20, \url{https://doi.org/10.1038/nmat1804}  


\bibitem{10} Donna C. Arnold,  Kevin S. Knight,  Finlay D. Morrison,  and Philip Lightfoot(2009) Ferroelectric-Paraelectric Transition in BiFeO$_3$ : Crystal Structure of the Orthorhombic $\beta$ Phase. PRL $\bf{102}$, 027602 \url{https://doi.org/10.1103/PhysRevLett.102.027602}


\bibitem{11} Donna C. Arnold, Kevin S. Knight, Gustau Catalan, Simon A. T. Redfern, James F. Scott, Philip Lightfoot, and Finlay D. Morrison(2010).The $\beta$ to $\gamma$ Transition in BiFeO$_3$ : A Powder Neutron Diffraction Study. Adv. Funct. Mater. $\bf{20}$, 2116-2123, \url{https://doi.org/10.1002/adfm.201000118}. 


\bibitem{12} M. Brahlek, A. K. Choquette, C. R. Smith, R. Engel-Herbert, and S. J. May(2017).Structural refinement of Pbnm-type perovskite films from analysis of half-order
diffraction peaks. Journal of Applied Physics $\bf{121}$, 045303; \url{https://doi: 10.1063/1.4974362}.

\bibitem{13}J. Enkovaara, C. Rostgaard, J. Mortensen, J. Chen, M. Dułak, L. Ferrighi, J. Gavnholt, C. Glinsvad, V. Haikola, H. Hansen, H. Kristoffersen, M. Kuisma, A. Larsen, L. Lehtovaara, M. Ljungberg, O. Lopez-Acevedo, P. Moses, J. Ojanen,T. Olsen, V. Petzold, N. Romero, J. Stausholm-Mller, M. Strange, G. Tritsaris, M. Vanin, M. Walter, B. Hammer, H. H$\ddot{a}$kkinen, G. Madsen, R. Nieminen, J. Nrskov, M. Puska, T. Rantala, J. Schitz, K. Thygesen, K. Jacobsen, Electronic structure calculations with GPAW: a real-space implementation of the projector augmented-wave method, J. Phys. Condens. Matter 22 (2010) 253202, \url{https://doi.org/10.1088/0953-8984/22/25/253202}.
\bibitem{14} J. Mortensen, L. Hansen, K. Jacobsen, Real-space grid implementation of the projector augmented wave method, Phys. Rev. B $\bf{71}$ (2005), 035109, \url{https://doi.org/10.1103/PhysRevB.71.035109}.
\bibitem{15} W. Kohn, L. Sham, Self-consistent equations including exchange and correlation effects, Phys. Rev. 140 (1965) A1133eA1138, \url{https://doi.org/10.1103/
PhysRev.140.A1133}.

\bibitem{16} P. Bl$\ddot{o}$chl, Projector augmented-wave method, Phys. Rev. B $\bf{50}$ (1994) 17953, \url{https://doi.org/10.1103/PhysRevB.50.17953}.

\bibitem{17} H. Monkhorst, J. Pack, Special points for Brillouin-zone integrations, Phys. Rev.
B $\bf{13}$ (1976) 5188, \url{https://doi.org/10.1103/PhysRevB.13.5188}.

\bibitem{18} G. Kresse, D. Joubert, From ultrasoft pseudopotentials to the projector augmented-wave method, Phys. Rev. B 59 (1999) 1758, \url{https://doi.org/
10.1103/PhysRevB.59.1758}.
\bibitem{19} H. Schlegel, Optimization of equilibrium geometries and transition structures, J. Comput. Chem. 3 (1982) 214, \url{https://doi.org/10.1002/jcc.540030212}.

\bibitem{20} P. Feynman, Forces in molecules, Phys. Rev. 56 (1939) 340, \url{https://doi.org/
10.1103/PhysRev.56.340}.


\bibitem{21} O. Nielsen, R. Martin, Quantum-mechanical theory of stress and force, Phys. Rev. B $\bf{32}$ (1985) 3780, \url{https://doi.org/10.1103/PhysRevB.32.3780}.

\bibitem{22} R. Wentzcovitch, J. Martins, First principles molecular dynamics of Li: test of a new algorithm, Solid State Commun. $\bf{78}$ (1991) 831, \url{https://doi.org/10.1016/
0038-1098(91)90629-A}.

\bibitem{23}J. Perdew, K. Burke, M. Ernzerhof, Generalized gradient approximation made simple, Phys. Rev. Lett. 77 (1996) 3865, \url{https://doi.org/10.1103/
PhysRevLett.77.3865}.

\bibitem{24} Vladimir I. Anisimov, Jan Zaanen, and Ole K. Andersen(1991), Band theory and Mott insulators:
Hubbard U instead of Stoner I, Phys. Rev. B $\bf{44}$, 943  \url{https://doi.org/10.1103/PhysRevB.44.943}

\bibitem{25} J. Tauc, R. Grigorovici, A. Vancu
Optical properties and electronic structure of amorphous germanium Phys. Status Solidi, $\bf{15}$ (1966), pp. 627-637, \url{10.1002/pssb.19660150224}.

\bibitem{26} Chan-Ho Yang, Daisuke Kan, Ichiro Takeuchi, Valanoor Nagarajan  and Jan Seidel(2012). Doping BiFeO$_{3}$ : approaches and enhanced functionality. Phys. Chem. Chem. Phys., $\bf{14}$, 15953–15962, \url{http://dx.doi.org/10.1039/C2CP43082G}.


\bibitem{27}  Tulika Maitra and A. Taraphder(2003). Magnetic, orbital, and charge ordering in the electron-doped manganites. Phys. Rev. B $\bf{68}$, 174416, \url{https://doi.org/10.1103/PhysRevB.68.174416}.

\bibitem{28} SUN et al(2005) Ab initio investigations of optical properties of the high-pressure phases of ZnO, Phys. Rev. B $\bf{71}$, 125132. \url{http://dx.doi.org/10.1103/PhysRevB.71.125132}


\bibitem{29} T.E. Ada, K.N. Nigussa, L.D. Daja, The effect of non-centrosymmetricity on optical and electronic properties of BaHfO${_3}$ perovskite, Computational Condensed Matter,$\bf{26}$, 2021, \url{https://doi.org/10.1016/j.cocom.2020.e00524}.



\bibitem{30} T.Shen,  C.Hu, H. L.Dai, W. L.Yang, H. C. Liu, 
 X. L. Wei(2015), First principles study of structural, electronic and optical properties of BiFeO${_3}$ in ferroelectric and paraelectric phases. aterials Research Innovations, $\bf{19}$, \url{https://doi.org/10.1179/1432891714Z.0000000001176}.

\bibitem{31} A H MacDonald $\emph{et al}$ 1979 J. Phys. C: Solid State Phys. $\bf{12}$ 2991,\url{
https://doi.org/10.1088/0022-3719/12/15/008}


\bibitem{32} Zhi-Ling Hou, Hai-Feng Zhou, Ling-Bao Kong , Hai-Bo Jin , Xin Qi, Mao-Sheng Cao, Materials Letters $\bf{84}$ (2012) 110-113,  \url{http://dx.doi.org/10.1016/j.matlet.2012.06.050}

\bibitem{33} Y. Hinuma et al.  Computational Materials Science $\bf{128}$ (2017) 140-184  \url{http://dx.doi.org/10.1016/j.commatsci.2016.10.015}


\bibitem{34} Thomas Olsen (2016), Designing in-plane heterostructures of quantum spin Hall insulators from first principles: 1T$^{'}$-MoS$_{2}$ with adsorbates Phys. Rev. B $\bf{88}$, 235106, \url{http://dx.doi.org/10.1103/PhysRevB.94.235106}

\bibitem{35} E. Yablonovitch(1993), Photonic band-gap structures, J. Opt. Soc. Am. B $\bf{10}$,  \url{http://dx.doi.org/10.1364/JOSAB.10.000283}

\bibitem{36} R. D. Meade, A. M. Rappe, K. D. Brommer, and J. D. Joannopoulos(1993), Nature of the photonic band gap: some insights from a
field analysis, J. Opt. Soc. Am. B $\bf{10}$
\url{http://dx.doi.org/10.1364/JOSAB.10.000328}.

\bibitem{37} G. Kurizki, B. Sherman, and A. Kadyshevitch(1993), Quantum electrodynamics in photonic band gaps: atomic-beam interaction with a defect mode, J. Opt. Soc. Am. B $\bf{10}$
\url{http://dx.doi.org/10.1364/JOSAB.10.000346}.


\bibitem{38}T. W Mossberg and M. Lewenstein(1993), Radiative properties of strongly driven atoms in the
presence of photonic bands and gaps, J. Opt. Soc. Am. B $\bf{10}$ \url{http://dx.doi.org/10.1364/JOSAB.10.000340}.


\bibitem{39} TD. R. Smith, R. Dalichaouch, N. Kroll, and S. Schultz(1993), Photonic band structure and defects in one and two dimensions, J. Opt. Soc. Am. B $\bf{10}$
\url{http://dx.doi.org/10.1364/JOSAB.10.000314}.

\bibitem{40}E. Yablonovitch and T. J. Gmitter(1990), Photonic band structure and defects in one and
two dimensions, J. Opt. Soc. Am. B $\bf{7}$
\url{http://dx.doi.org/10.1364/JOSAB.7.0001792}.

\bibitem{41} Cebela, M., Ceramics International (2016),\url{ http://dx.doi.org/10.1016/j.ceramint.2016.10.074}

 

\bibitem{42} Lebeugle $\emph{et al}$ (2008), Electric-Field-Induced Spin Flop in BiFeO$_{3}$ Single Crystals at Room Temperature, Phys. Rev. B $\bf{100}$, 227602.\url{ http://dx.doi.org/10.1103/PhysRevLett.100.227602}




\end{thebibliography}

\end{document}